\input{aipcheck}

\documentclass[
    ,final            
  ]
  {aipproc}

\layoutstyle{6x9}

\begin{document}

\title{Hadron Structure with Dimuon Production}



\author{J.C.~Peng}{
  address={University of Illinois at Urbana-Champaign, Urbana, IL 61801}
}


\begin{abstract}
Dimuon production has been studied in a series of fixed-target
experiments at Fermilab during the last two decades. Highlights
from these experiments, together wih recent
results from the Fermilab E866 experiment, are presented. Future
prospects for studying the parton distributions in the nucleons
and nuclei using dimuon production at Fermilab and J-PARC  
are also discussed.
\end{abstract}

\maketitle


\section{Introduction}

While our current knowledge on the partonic substructures of
nucleons and nuclei is obtained mostly from deep-inelastic
scattering experiments, the dimuon production experiments
also play an important and unique role in providing complementary
information. During the last two decades, a series of fixed-target dimuon
production experiments (E772, E789, E866) have been carried out 
using 800 GeV/c proton beam at Fermilab. 
At 800 GeV/c, the dimuon data contain Drell-Yan continuum up to 
dimuon mass of $\sim 15$ GeV as well as quarkonium productions
(J/$\Psi$, $\Psi^\prime$, and $\Upsilon$ resonances). The Drell-Yan
process and quarkonium productions often provide complementary
information, since Drell-Yan is an electromagnetic process
via quark-antiquark annihilation while the quarkonium production
is a strong interaction process dominated by gluon-gluon fusion
at this beam energy.

The Fermilab dimuon experiments covers a broad range of physics
topics. The Drell-Yan data have provided informations on the 
antiquark distributions in the nucleons~\cite{pat92,hawker98,
peng98,towell01} and nuclei~\cite{alde90,vasiliev99}. 
These results showed the surprising results that the
antiquark distributions in the
nuclei are not enhanced~\cite{alde90,vasiliev99}, 
contrary to the predictions of models which explain the
EMC effect in term of nuclear
enhancement of exchanged mesons. Moreover, the Drell-Yan cross 
section ratios $p+d/p+p$ clearly establish the
flavor asymmetry of the $\bar d$ and $\bar u$ distributions in
the proton, and they map out the $x$-dependence of this 
asymmetry~\cite{hawker98,peng98,towell01}.
Pronounced nuclear dependences of quarkonium productions have
been observed for J/$\Psi$, $\Psi^\prime$, and $\Upsilon$ resonances
~\cite{alde91a,alde91b,kowitt94,leitch95}. It was found that these
nuclear effects scale with the kinematic variable $x_F$ rather than
$x_2$ (the Bjorken-$x$ of the parton in the nulceus), suggesting
the importance of initial- and final-state interactions. A striking 
behavior of the nuclear dependence as a function of $p_T$, reminiscent
of the Cronin effect, was also observed. The nuclear Drell-Yan cross 
sections also exhibit $x_F$ as well as $p_T$ dependences~\cite{vasiliev99,
pat99,johnson01,johnson02}, which are weaker than the quarkonium 
nuclear dependences but can provide information on
the energy loss of quarks traversing the nucleus~\cite{garvey03}.
The differential cross sections for Drell-Yan~\cite{pat94}, 
charmonium~\cite{schub96}, and
bottomonium~\cite{zhu08} productions have also been reported.
In addition, the decay angular distributions for Drell-Yan~\cite{pat99,
zhu07}, J/$\Psi$~\cite{chang03}, and $\Upsilon$ resonances~\cite{brown01}
have been measured. Information on $D$ and $B$ meson productions
and decays has also been extracted through either the dimuon 
measurement~\cite{mishra94,jansen95,pripstein00}
or the open-aperture dihadron measurement~\cite{leitch94}.
Several review articles covering some of these results are
available~\cite{pat99,garvey01,reimer07}. In this article, we
will focus on the recent results from experiment E866 and 
future prospect of dimuon experiments at Fermilab and
J-PARC.

\section{Recent Results from E866}

\subsection{Angular distributions of proton-induced Drell-Yan}

Despite the success of perturbative QCD in describing the
Drell-Yan cross
sections, it remains a challenge to understand the angular
distributions of the Drell-Yan process. Assuming dominance of the
single-photon process, a general expression for the Drell-Yan
angular distribution is~\cite{lam78}
\begin{equation}
\frac {d\sigma} {d\Omega} \propto 1+\lambda \cos^2\theta +\mu \sin2\theta
\cos \phi + \frac {\nu}{2} \sin^2\theta \cos 2\phi,
\label{eq:eq1}
\end{equation}
\noindent where $\theta$ and $\phi$ denote the polar and azimuthal angle,
respectively, of the $l^+$ in the dilepton rest frame. In the ``naive"
Drell-Yan model, where the transverse momentum of the quark is ignored
and no gluon emission is considered, $\lambda =1$ and $\mu = \nu =0$ are
obtained. QCD effects~\cite{chiappetta86} and
non-zero intrinsic transverse momentum of the quarks~\cite{cleymans81}
can both lead to $\lambda \ne 1$ and $\mu, \nu \ne 0$. However,
$\lambda$ and $\nu$ should still
satisfy the relation
$1-\lambda = 2 \nu$~\cite{lam78}. This so-called Lam-Tung relation, obtained as
a consequence of the spin-1/2 nature of the quarks, is analogous
to the Callan-Gross relation~\cite{callan69}
in deep-inelastic scattering. 

The first measurement of the Drell-Yan angular distribution was
performed by the NA10 Collaboration for $\pi^- + W$ with the 
highest statistics at 194
GeV/c~\cite{falciano86,guanziroli88}.
The $\cos 2 \phi$ angular dependences showed a sizable $\nu$,
increasing with dimuon transverse momentum ($p_T$) and reaching a value
of $\approx 0.3$ at $p_T = 2.5$ GeV/c. The Fermilab E615 Collaboration
subsequently performed a measurement of $\pi^- + W$ Drell-Yan production
at 252 GeV/c with broad coverage in the
decay angle $\theta$~\cite{conway89}. The E615 data showed that
the Lam-Tung relation, $2\nu = 1 - \lambda$, is clearly violated.

The NA10 and E615 results on the Drell-Yan angular distributions strongly
suggest that new effects beyond conventional perturbative QCD are present.
Brandenburg, Nachtmann and Mirke suggested that a factorization-breaking
QCD vacuum may lead to a
correlation between the transverse spin of the antiquark in the pion and
that of the quark in the nucleon~\cite{brandenburg93}.
This would result in a non-zero
$\cos 2\phi$ angular dependence consistent with the data. 
Several authors
have also considered higher-twist effects from quark-antiquark binding
in pions~\cite{brandenburg94,eskola94}. 
However, the model is strictly applicable
only in the $x_\pi \to 1$ region, while the NA10 and E615 data exhibit
nonperturbative effects over a much broader kinematic region.

\begin{figure}[tb]
\includegraphics*[height=.55\textheight]{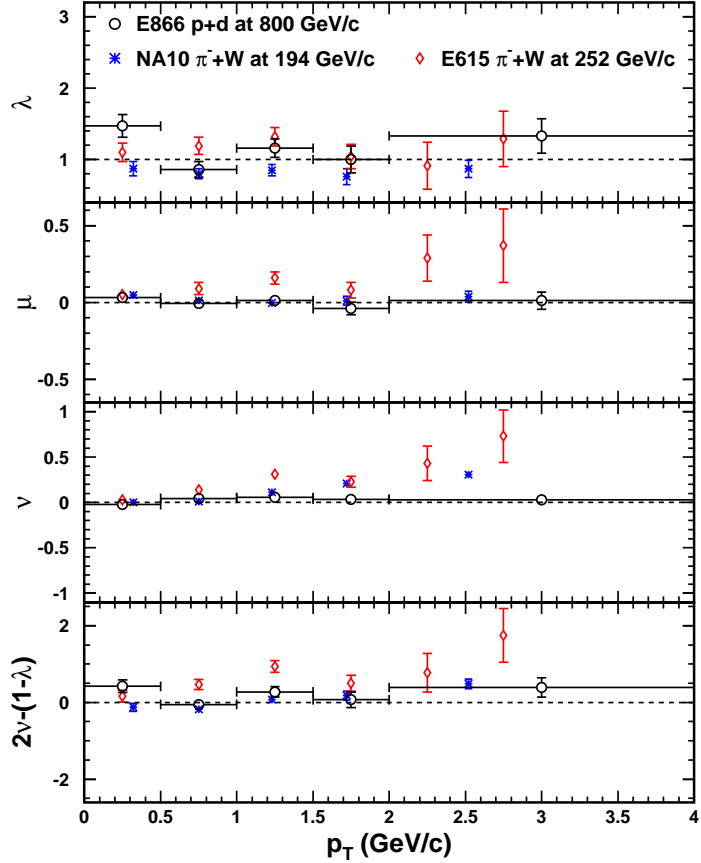}
\caption{ Parameters $\lambda, \mu, \nu$ and $2\nu - (1-\lambda)$
vs.\ $p_T$ in the Collins-Soper frame. Solid circles are for
E866 $p+d$ at 800 GeV/c, crosses are for NA10 $\pi^- + W$ at 194 GeV/c, and
diamonds are E615 $\pi^- + W$ at 252 GeV/c. The error bars
include the statistical uncertainties only.}
\label{pdfig1}
\end{figure}

More recently, Boer pointed out~\cite{boer99}
that the $\cos 2 \phi$ angular dependences
observed in NA10 and E615 could be due to the $k_T$-dependent
parton distribution function $h_1^\perp$. This so-called Boer-Mulders
function~\cite{boer98} is an example of a novel type of $k_T$-dependent
parton distribution function, and it
characterizes the correlation of a quark's transverse spin and
its transverse momentum, $k_T$, in an unpolarized nucleon. 

To shed additional light on the origins of the NA10 and E615 Drell-Yan
angular distributions, we have analyzed $p+p$ and $p+d$ Drell-Yan angular
distribution data at 800 GeV/c from Fermilab E866. 
There has been no report on
the azimuthal angular distributions for proton-induced Drell-Yan -- all
measurements so far have been for polar angular
distributions~\cite{pat99,chang03,brown01}.
Moreover, proton-induced Drell-Yan data provide a test of
theoretical models. For example, the $\cos 2\phi$ dependence is expected
to be much reduced in proton-induced Drell-Yan if the underlying mechanism
involves the Boer-Mulders functions. This is due to the expectation that
the Boer-Mulders functions are small for the sea-quarks. However, if the
QCD vacuum effect~\cite{brandenburg93} is the origin of
the $\cos 2 \phi$ angular dependence, then the azimuthal behavior of
proton-induced Drell-Yan should be similar to that of pion-induced
Drell-Yan. 
Finally, the validity of the Lam-Tung relation has never been
tested for proton-induced Drell-Yan, and the present
study provides a first test.

The Fermilab E866 experiment was performed using the upgraded Meson-East
magnetic pair spectrometer. Details of the experimental setup have been
described elsewhere~\cite{hawker98,peng98,towell01}. An 800 GeV/c 
primary proton
beam with up to $2 \times 10^{12}$ protons per 20-second beam spill
was incident upon 50.8 cm long cylindrical
stainless steel target flask containing liquid deuterium.
The detector system consisted of
four tracking stations and a momentum analyzing magnet.
From the momenta of the $\mu^+$ and $\mu^-$, kinematic variables of
the dimuons ($x_F, m_{\mu\mu}, p_T$) were readily reconstructed.
The muon angles $\theta$ and $\phi$ in the Collins-Soper
frame~\cite{collins77} were also calculated. To remove the
quarkonium background,
only events with $4.5 <m_{\mu\mu}<
9$ GeV/c$^2$ or $m_{\mu\mu} > 10.7$ GeV/c$^2$ were analyzed.
A total of $\sim$54,000 $p+p$ and $\sim$118,000
$p+d$ Drell-Yan events covering the decay angular range $-0.5 < \cos\theta
<0.5$ and $-\pi < \phi < \pi$ remain. 

\begin{figure}[tb]
\includegraphics*[width=\linewidth]{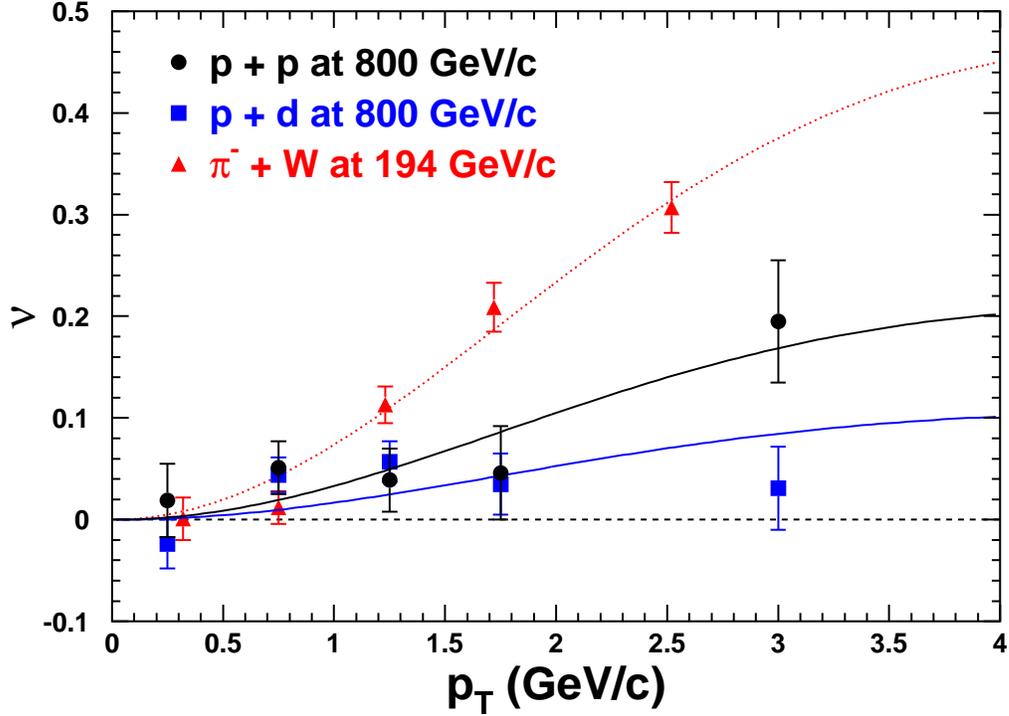}
\caption{Parameter $\nu$ vs.\ $p_T$ in the Collins-Soper
frame for several Drell-Yan
measurements. Fits to the data using Eq. 3 and $M_C=2.4$ GeV/c$^2$ are
also shown.}
\label{ppfig2}
\end{figure}

Figure 1 shows the results~\cite{zhu07} on the angular 
distribution parameters $\lambda, \mu,$ and
$\nu$ vs.\ $p_T$. To extract these parameters, the Drell-Yan data were
grouped into 5 bins in $\cos\theta$ and 8 bins in $\phi$ for each $p_T$
bin. A least-squares fit
to the data using Eq. 1 to describe the angular distribution was
performed. Only statistical errors are shown
in Fig. 1.
For comparison with the $p+d$ Drell-Yan data, the NA10 $\pi^- + W$ data
at 194 GeV/c and the E615 $\pi^- + W$ data at 252 GeV/c are also shown
in Fig. 1. To test the validity of the Lam-Tung relation, also shown
in Fig. 1 is the quantity, $2\nu - (1-\lambda)$, for all three
experiments. For $p+d$ at 800 GeV/c, Fig. 1 shows that $\lambda$ is
consistent with 1, in agreement with previous 
studies~\cite{pat99,chang03,brown01},
while $\mu$ and $\nu$ deviate only slightly from zero.
This is in contrast to the pion-induced Drell-Yan
results, in which much larger values of
$\nu$ are found. It is also interesting to note that while E615 clearly
establishes the violation of the Lam-Tung relation, the NA10 and the
$p+d$ data are largely consistent with the Lam-Tung relation.

The $p+d$ results put constraints on theoretical models that
predict large $\cos 2 \phi$ dependence originating from QCD vacuum effects.
They also suggest that the Boer-Mulders function
$h_1^\perp$ for sea quarks is significantly smaller than for
valence quarks. A recent analysis~\cite{zhang08} of
the $p+d$ $\cos 2 \phi$ data
showed that the sea-quark Boer-Mulders functions are indeed smaller by a factor
$\sim 5$ than the valence-quark Boer-Mulders functions.
This analysis also indicated that the E866 $p+d$
data are consistent with the $u$ and $d$ Boer-Mulders
functions having the same signs. However, the
$p+d$ data alone can not provide an unambiguous determination of the
flavor dependence of the Boer-Mulders functions. A comparison of the 
$p+p$ and $p+d$ data would further constrain the flavor dependence 
of the Boer-Mulders functions.

Figure 2 shows the preliminary result of $\nu$ for $p+p$ reaction at 
800 GeV/c. Also shown in Fig. 2 are the E866 $p+d$ and the NA10 
$\pi^- +W$ data at 194 GeV/c. The data are fitted with the expression
suggested by Boer~\cite{boer99}:
\begin{eqnarray}
\nu = 16 \kappa_1 \frac {p_T^2 M_C^2} {(p_T^2 + 4 M_C^2)^2},
\label{eq:eq3}
\end{eqnarray}
\noindent where $\kappa_1$ is proportional to the product of the
$h_1^\perp$ functions for the projectile and the target, and $M_C$ is
a constant fitting parameter. Boer obtained $\kappa_1 = 0.47 \pm 0.14$
and $M_C = 2.4 \pm 0.5$ GeV/c$^2$ for fitting the NA10 data, as
shown in Fig. 2. A fit to the E866 $p+p$ and $p+d$ data for
$M_C = 2.4$ GeV/$c^2$ yields $\kappa_1 = 0.21 \pm 0.055$ and
$\kappa_1 = 0.11 \pm 0.04$, respectively.
These data should provide additional information on the origins of
the azimuthal angular dependence of Drell-Yan, as well as the flavor
dependence of the Boer-Mulders functions.

\subsection{ $\Upsilon$ production for $p+p$ and $p+d$ interactions}

In the CERN NA51~\cite{na51a} and Fermilab E866~\cite{hawker98,
peng98,towell01}
experiments on proton-induced dimuon production, a striking difference was
observed for the Drell-Yan cross sections between $p+p$ and $p+d$.
As the underlying mechanism for the Drell-Yan process involves
quark-antiquark annihilation, this difference has been attributed to
the asymmetry between the up and down sea quark distributions
in the proton. From the $\sigma (p+d)_{DY}/2\sigma(p+p)_{DY}$ ratios
the Bjorken-$x$ dependence of the sea-quark
$\bar d / \bar u$ flavor
asymmetry has been extracted~\cite{hawker98,peng98,towell01,na51a}.

\begin{figure}[tb]
\includegraphics*[width=\linewidth]{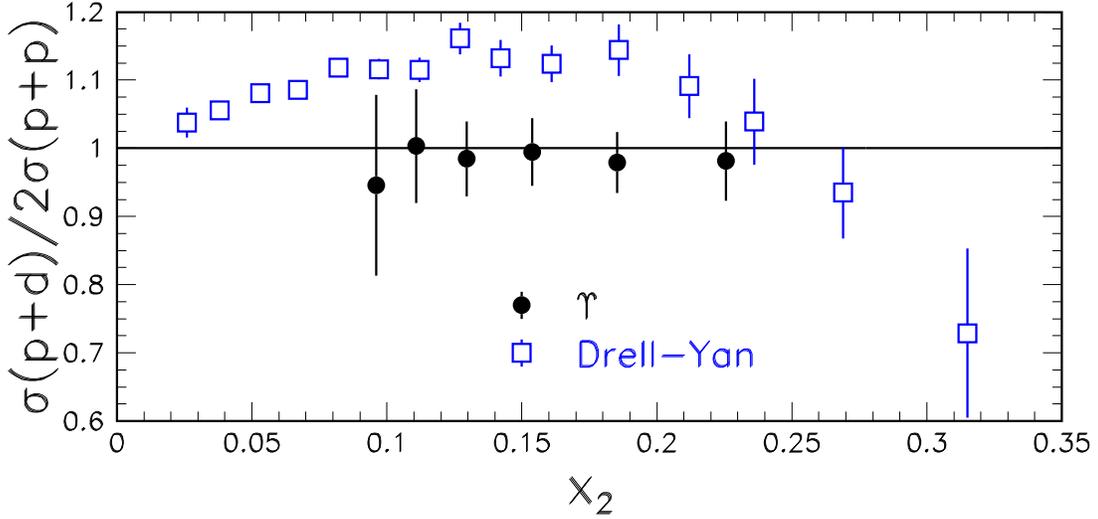}
\caption{The E866 $\sigma(p + d) /2 \sigma(p + p)$ cross section
ratios for $\Upsilon$ resonances as a function of $x_2$. The
corresponding ratios for Drell-Yan cross sections
are also shown. The error bars are statistical only.}
\label{crosfig3}
\end{figure}

The Fermilab E866 dimuon experiment also recorded a large number of
$\Upsilon \to \mu^+ \mu^-$ events. Unlike the electromagnetic
Drell-Yan process, quarkonium
production is a strong interaction dominated by the subprocess of
gluon-gluon fusion at this beam energy~\cite{jansen,vogt99}. Therefore, the
quarkonium production cross
sections are primarily sensitive to the gluon distributions in the
colliding hadrons. The $\Upsilon$ production ratio, $\sigma(p+d \to \Upsilon)
/2 \sigma(p+p \to \Upsilon)$, is expected to
probe the gluon content in the neutron relative to that in the
proton~\cite{piller}. While it is generally assumed that the gluon
distributions in the proton and neutron are identical, this assumption
is not based on any fundamental symmetry and has not been tested
experimentally. A possible mechanism for generating different gluon
distributions in the proton and neutron, as pointed out by Piller
and Thomas~\cite{piller}, is the violation of charge symmetry in the quark
and antiquark distributions in the nucleons. A precise measurement of the
$\sigma(p+d \to \Upsilon)/2 \sigma(p+p \to \Upsilon)$ ratios
would provide a constraint on the asymmetry
of gluon distribution in the proton versus that in the neutron.

The $\sigma (p+d)/2\sigma(p+p)$ ratios for $\Upsilon (1S+2S+3S)$ production
are shown in Fig.\  3 as a function of $x_2$. Most of the 
systematic errors cancel for these
ratios, with a remaining $\approx$ 1\% error from the rate dependence
and target compositions~\cite{towell01}.
Figure 3 shows that these ratios are consistent with unity, in striking
contrast to the corresponding values~\cite{towell01} for the Drell-Yan process,
also shown in Fig.\  3.
The difference between the Drell-Yan and the
$\Upsilon$ cross section ratios clearly reflect the different
underlying mechanisms in these two processes. The Drell-Yan process,
dominated by the $u - \bar u$ annihilation subprocess, leads to the
relation $\sigma (p + d)_{DY}/2 \sigma (p+p)_{DY} \approx \frac {1}{2}
(1 + \bar u_n(x_2)/\bar u_p (x_2)) = \frac {1}{2}
(1 + \bar d_p(x_2)/\bar u_p(x_2))$, where $\bar q_{p,n}$ refers to
the $\bar q$ distribution in the proton and neutron, respectively.
For $\Upsilon$ production, the dominance of the
gluon-gluon fusion subprocess at this beam energy
implies that $\sigma (p+d \to \Upsilon)/ 2 \sigma(p+p \to \Upsilon)
\approx \frac {1}{2} (1+g_n(x_2)/g_p(x_2))$. Figure 3 shows that
the gluon distributions in the proton ($g_p$) and neutron ($g_n$)
are very similar over the $x_2$ range $0.09 < x_2 < 0.25$. The overall
$\sigma (p+d \to \Upsilon)/ 2 \sigma(p+p \to \Upsilon)$ ratio, integrated
over the measured kinematic range, is $0.984 \pm 0.026 (\rm{stat.})
\pm 0.01 (\rm{syst.})$. 
The $\Upsilon$ data indicate that the gluon distributions in the
proton and neutron are very similar.
These results are consistent with no
charge symmetry breaking effect in the gluon distributions. 

\section{Future Prospects at Frmilab and J-PARC}

Future fixed-target dimuon experiments have been proposed at
the 120 GeV Fermilab Main Injector (FMI) and the 50 GeV J-PARC
facilities. The Fermilab proposal~\cite{e906}, E906, has been 
approved and is expected to start data-taking around 2011. Two
dimuon proposals (P04~\cite{p04} and P24~\cite{p24}) have also
been submitted to the J-PARC for approval. The lower beam energies
at FMI and J-PARC present opportunities for extending the $\bar d/\bar u$
and the nuclear antiquark distribution measurements to larger 
$x$ ($x>0.25$). For given values of $x_1$ and $x_2$, the Drell-Yan cross 
section is proportional to $1/s$, hence a gain of $\sim 16$ times in the
Drell-Yan cross sections can be obtained at the J-PARC energy of 50 GeV.
Since the perturbative process gives a symmetric $\bar d/ \bar u$ while
non-perturbative processes are necessary to generate an asymmetric
$\bar d/ \bar u$ sea, it would be very important to extend the Drell-Yan
measurements to kinematic regimes beyond the current limits. Another
advantage of lower beam energies is that a much more sensitive study of the
partonic energy loss in nuclei could be carried out using the Drell-Yan
nuclear dependence~\cite{garvey03}.

The dimuon physics program at J-PARC is proposed to be carried out in
several stages. Since 30 GeV proton beam will be available at the initial
phase of J-PARC, the first measurements will focus on $J/\Psi$ production
at 30 GeV. This will be followed by measurements of Drell-Yan and
quarkonium production at 50 GeV after the beam energy is upgraded to
50 GeV. Experiments using polarized target could already be performed
with unpolarized beams. When polarized proton beam becomes available at
J-PARC, a rich and unique program on spin physics could also be pursued
at J-PARC using the dimuon spectrometer.

An important feature of $J/\Psi$ production using 30 or 50 GeV proton
beam is the dominance of the quark-antiquark annihilation subprocess.
This is in striking contrast to $J/\Psi$ production at 800 GeV (Fermilan
E866) or at 120 GeV (Fermilab E906), where the gluon-gluon fusion is the
dominant process. This suggests an exciting opportunity to use $J/\Psi$
production at J-PARC as an alternative method to probe antiquark
distribution. 

With the possibility to accelerate polarized proton beams at J-PARC,
the spin structure of the proton can also be investigated with the
proposed dimuon experiments. In particular, polarzied Drell-Yan process
with polarized beam and/or polarized target at J-PARC would allow a unique
program on spin physics complementary to polarized DIS experiments
and the RHIC-Spin program. Specific physics topics include the measurements
of T-odd Boer-Mulders distribution function in unpolarized Drell-Yan,
the extraction of T-odd Sivers distribution functions in singly
transversely polarized Drell-Yan, the helicity distribution of
antiqaurks in doubly longitudinally polarized Drell-Yan, and the
transversity distribution in doubly transversely polarized
Drell-Yan. It is worth
noting that polarized Drell-Yan is one of the major physics program at the
GSI Polarized Antiproton Experiment (PAX). The RHIC-Spin program will
likely provide the first results on polarized Drell-Yan. However, the
high luminosity and the broad kinematic coverage for the
large-$x$ region at J-PARC would allow some unique measurements to be
performed in the J-PARC dimuon experiments.

I am very grateful to my many collaborators on the E772, E789, and
E866 experiments at Fermilab. I would also acknowledge the collaboration
with Dr. Shin'ya Sawada and Dr. Yuji Goto on the J-PARC dimuon proposals.


\end{document}